\documentclass[twocolumn,showpacs,prl]{revtex4}

\usepackage{graphicx}  
\usepackage{dcolumn}
\usepackage{amsmath}
\usepackage{amssymb}%
\begin{document}

\preprint{}

\title{Velocity correlations in dense granular flows}
 \author{O. Pouliquen}
\affiliation{IUSTI, Universit\'e de Provence - CNRS, 5 rue Enrico Fermi, 13453 Marseille cedex 13, France}

\date{\today}
%\vspace{20mm}

\begin{abstract}
Velocity fluctuations of grains flowing down a rough inclined plane are experimentally studied.  The grains at the free surface exhibit fluctuating motions, which are correlated over few grains diameters. The characteristic correlation length is shown to depend on the inclination of the plane and not on the thickness of the flowing layer.  This result strongly supports the idea that dense granular flows are controlled by a characteristic length larger than the particle diameter.
 \end{abstract}
\maketitle
%\vspace{1.5cm}
 Dry cohesionless granular material can flow like a liquid, for example in an hourglass or on an inclined plane. However, the flow properties are not yet well understood and constitutive equations appropriate to describe the dense flow regime  when particles are in close contact are still matter of debate \cite{gdrmidi04,pouliquen02}. One of the difficulty is that, in the dense  regime, particles do not interact through  binary collisions like in dilute and agitated granular gases  \cite{jenkins83,goldhirsch99}. They rather experience contact with several neighboring particles at the same time. The description of the multi-particles interactions is important in the development of constitutive equations. Recent theoretical models try to take into account the existence of this contact network through different approaches. Some models describe the flow as a mixture of solid and liquid phases \cite{volson03} or try to account for the frictional contacts by  adding a frictional term in collisional theories \cite{johnson87,louge03}. Other models consider the presence of arches  in the flow  \cite{mills99,mills00} or account for some non local propagation of momentum through the contact network \cite{rajchenbach03,pouliquen01}.  In this paper we will more specifically refer to a recent model proposed by Ertas and Halsey \cite{ertas02} to describe granular flows on an inclined plane. By introducing the idea of correlated motion of grains in clusters-like structures, they have been able by simple scaling arguments to recover some important results observed for flows down inclined plane.  The experimental study presented in this paper is inspired by their approach, trying to evidence the existence of correlated motions in granular flows down inclined planes. 
 
\begin{figure}[!ht]
  \begin{center}
\includegraphics[scale=0.17]{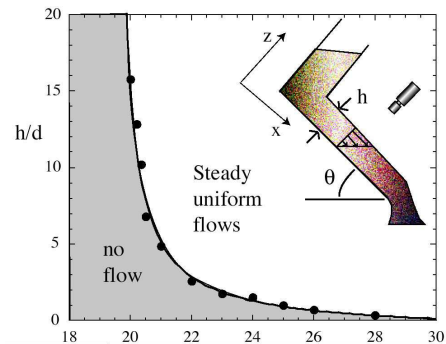}
  \caption{Flow threshold:  ($\bullet$) experimental measurements of the deposit $h_{stop}(\theta)$. Inset: sketch of the experimental set up }
\label{fig1}
  \end{center}
 \end{figure}

The inclined plane configuration is obtained when a granular layer of thickness $h$ flows down a rough surface inclined at an angle $\theta$ (Fig. \ref{fig1}). Experiments \cite{pouliquen99,gdrmidi04} and numerical simulations \cite{silbert01,silbert03} have revealed two important results. First, for a given inclination $\theta$, a minimum thickness $h_{stop}(\theta)$ exists below which no flow is possible. This thickness is evidenced by the deposit remaining on the plane once the flow stops.  Second, the variation of  the depth averaged velocity $u$  with the inclination $\theta$ and thickness $h$ appears to be linked to the deposit function $h_{stop}(\theta)$ through a flow rule: 
\begin{equation}
\frac{u}{\sqrt{gh}}= \beta \frac{h}{h_{stop}(\theta)} ,
\label{flowrule}
\end{equation}
where $g$ is gravity and $\beta$ is a constant equal to 0.13 for spheres. 
 The fact that the influence of $\theta$ on the flow velocity comes into play through the deposit function $h_{stop}(\theta)$  is not straightforward and could be simply the result of a coincidence \cite{louge03}. However,  an interesting interpretation of relation \ref{flowrule} has been proposed by Ertas and Halsey \cite{ertas02}. Their idea is that motion of grains in dense granular flows occurs through clusters, which size is controlled by the stress distribution. In this picture, it is clear that no flow would be possible for layers thinner than the characteristic size of the clusters. This immediately gives an interpretation of the deposits: the deposit thickness $h_{stop}$ would be an indirect measurement of the cluster size. Based on the existence of a characteristic cluster size, Ertas and Halsey \cite{ertas02} have developed  scaling arguments similar to  a Prandtl mixing length turbulent theory and are able to derive the flowing rule \ref{flowrule} \cite{gdrmidi04}. 
 
  This heuristic approach is appealing because it gives simple interpretations for the deposit thickness and the flow rule observed in inclined plane, but also because it gives new insights for the granular flow rheology. As discussed in \cite{gdrmidi04}, this picture of correlated grains motion gives a way to  link the different flow regimes observed with granular materials:  the slow quasi-static regimes would correspond to strongly correlated motions whereas the kinetic collisional regime would correspond to a regime where spatial correlations vanish. To our knowledge, no experimental evidence exists to support this description.   
  
  Whereas correlations are studied in details in other disordered systems like glasses \cite{bennemann99,berthier04} or foams \cite{durian95,debregeas02,ono03}, only few studies address the question of correlated motions in granular flows. Most of them concern the slow quasi-static regime where long range correlations have been observed but no characteristic length  has been evidenced \cite{kuhn99,radjai03}. In the dense flow regime, Bonamy et al \cite{bonamy02} have investigated correlated motions in a rotating drum.  Their analysis shows that clusters of different sizes exist but that no characteristic length scale emerges. 
The purpose of this paper is to experimentally study spatial correlations in the inclined plane configuration  and check  if  the deposit thickness observed can indeed be interpreted as an indirect measure of the correlation length as suggested in \cite{ertas02}. 

\begin{figure}[!ht]
  \begin{center}
\includegraphics[scale=0.55]{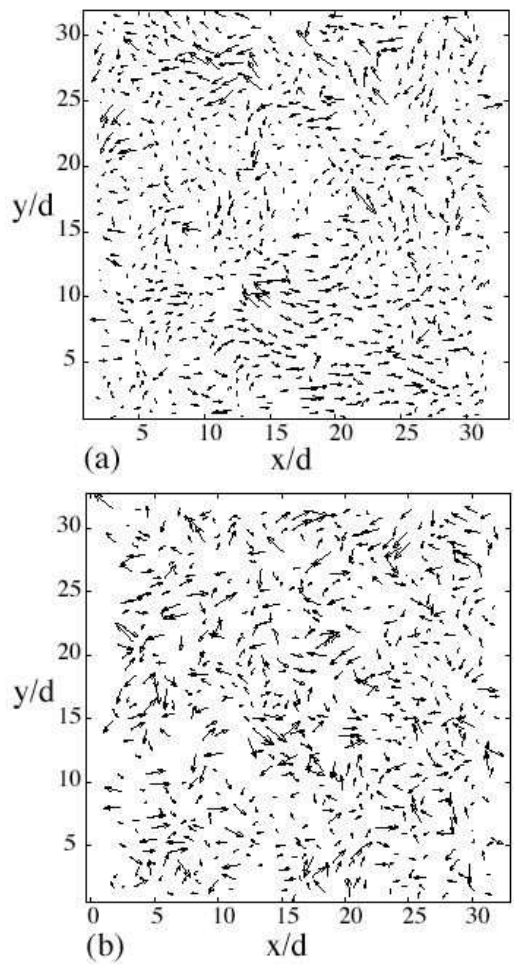}
  \caption{Fluctuating velocity of particles at the free surface (a) $\theta=21^o$, $h/d=13.7$;(b) $\theta=26^o$, h/d=9.3 . }
\label{fig2}
  \end{center}
 \end{figure}

\begin{figure}[!ht]
  \begin{center}
\includegraphics[scale=0.25]{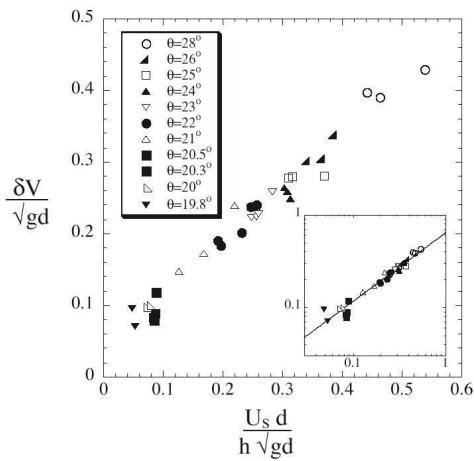}
  \caption{Mean velocity fluctuations $\frac{\delta V}{\sqrt{gd}}$  as a function of the mean shear rate $\frac{U_s d}{h\sqrt{gd}}$ for different inclinations $\theta$. Inset: log-log plot. }
\label{fig3}
  \end{center}
 \end{figure}

The experimental set up is presented in the inset of Fig. \ref{fig1} and is the same as the one used in \cite{pouliquen99}. A two meters long and 70 cm wide plane is made rough by gluing particles on it and can be inclined from horizontal at an angle $\theta$.  The material (glass beads 0.5 mm in diameter) flows from a hopper, the flow rate being controlled by the opening of the gate. The inclination $\theta$ and the thickness $h$ of the layer are the two control parameters. Steady uniform flows are obtained with this device in a finite range of inclination and thickness as shown in Fig. \ref{fig1}. The flow threshold delimited by the curve $h_{stop}(\theta)$, has been determined by measuring the deposit thickness once the flow stops.  In order to study correlated motions, we have chosen to precisely study the individual grains motion at the free surface using a high speed video camera placed above the plane. From movies recorded between 500 and 1000  frames per second, we  measure the velocity of the particles using a precise particle tracking algorithm. The mean free surface velocity $U_s \vec{e}_x$ is  computed by averaging over 1 second of flow and over all the detected particles. It is then subtracted to all individual particle velocities to get the instantaneous fluctuating velocity field. For each time $t$ of our movie, we then have the position $\left(x^i,y^i\right)$ of  particle $i$, and its velocity relative to the mean free surface motion $\left(u^i_x, u^i_y\right)$. Example of fluctuating velocity fields are presented in Figs. \ref{fig2} (a) and (b) for two different inclinations.  Fig. \ref{fig2} (a) obtained at low inclination clearly exhibits correlated motions. By contrast, the velocity field is much more disordered with less correlations at high inclination as shown in Fig. \ref{fig2} (b). 
Before studying spatial correlations, we first analyze the amplitude of the fluctuations by computing the mean fluctuating velocity $\delta V=(<(u_x^i)^2+(u_y^i)^2>)^{1/2}$ averaged over the particles and over 1second of flow.  In Fig.  \ref{fig3},  the dimensionless fluctuating velocity amplitude $\delta V/\sqrt{gd}$ is plotted as a function of the mean shear rate i.e. $U_s/h \sqrt{g/d}$.  All the data obtained for different inclinations and different flow thicknesses collapse on a single curve, showing that the fluctuations at the free surface are entirely driven by the shear rate. The plot seems to indicate an affine relation between fluctuations and shear rate. However,  if one tries to fit the data by a power law in order to compare with results published in the literature,  one get a power equal to 0.7 , compatible with measurements in other flow configurations  \cite{gdrmidi04}. 

\begin{figure}[!ht]
  \begin{center}
\includegraphics[scale=0.22]{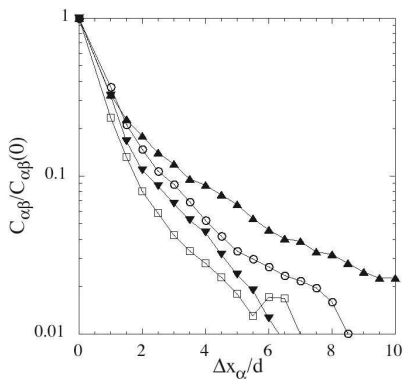}
  \caption{The four normalized correlation functions for $\theta=20^o$ and $h/d=19.8$; $C_{xx}$ ($\blacktriangle$)  , $C_{yy}$ ({\Large $\circ$}) , $C_{yx}$ ($\blacktriangledown$), $C_{xy}$   ( $\square$).}
\label{fig4}
  \end{center}
 \end{figure}

We now turn to the study of the spatial correlations. 
To this end, we defined four correlations function $C_{\alpha \beta}$ corresponding to the correlation of the $\beta$ velocity component along the $\alpha$ direction. For example,  the correlation $C_{xx}(\Delta x)$ at a distance $(\Delta x)$ is defined as follow:
$$
C_{xx}(\Delta x)=\frac{\sum_t \left(\sum_{i,j}  u^i_x u^j_x \delta(x^i+\Delta x-x^j) \delta(y^i-y^j)\right)}{\sum_t \left(\sum_{i,j} \delta(x^i+\Delta x-x^j) \delta(y^i-y^j)\right)},
$$
where the function $\delta(x)$ is a peaked function. We have chosen a Gaussian of width 0.4 particle diameter for $\delta$ but have checked that the results  are insensitive to this choice, as long as $\delta$ is narrow.  Fig. \ref{fig4} shows the four correlation functions for $\theta=20^o$ and $h/d=19.8$. The four functions have the same characteristics: a first rapid decrease, followed after $2d$ by a roughly exponential decrease over a characteristic length scale of few particles diameters. The function  $C_{xx}$ is larger than the others,  meaning that the correlations are stronger for the longitudinal velocity along the flow direction. In the following, we show  the influence of the inclination $\theta$ and flow thickness $h$ on  $C_{xx}$ only, but the same qualitative behaviors hold for the three other correlation functions. 

\begin{figure}[!ht]
  \begin{center}
\includegraphics[scale=0.35]{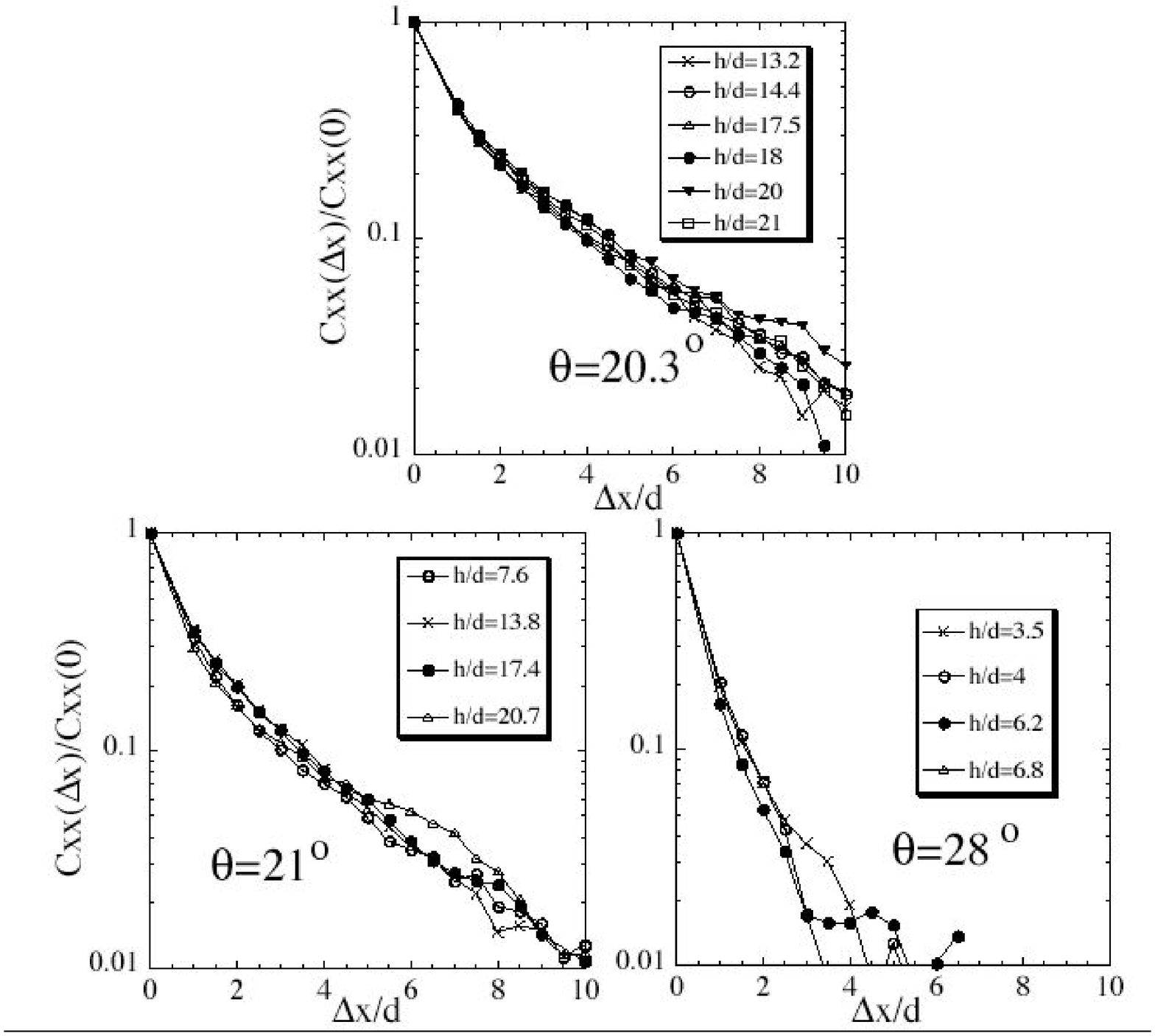}
  \caption{Normalized correlation function $C_{xx}$ for different flow thicknesses at three inclinations. }
\label{fig5}
  \end{center}
 \end{figure}

\begin{figure}[!ht]
  \begin{center}
\includegraphics[scale=0.30]{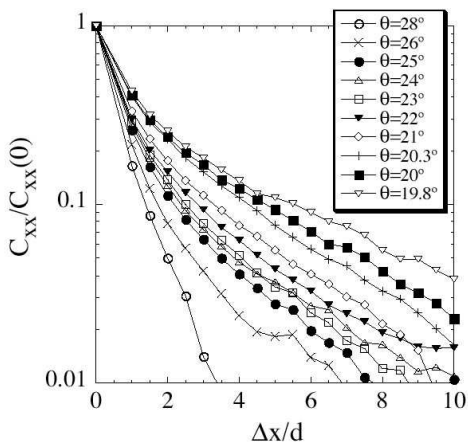}
  \caption{Mean normalized correlation function for different $\theta$.  }
\label{fig6}
  \end{center}
 \end{figure}

\begin{figure}[!ht]
  \begin{center}
\includegraphics[scale=0.30]{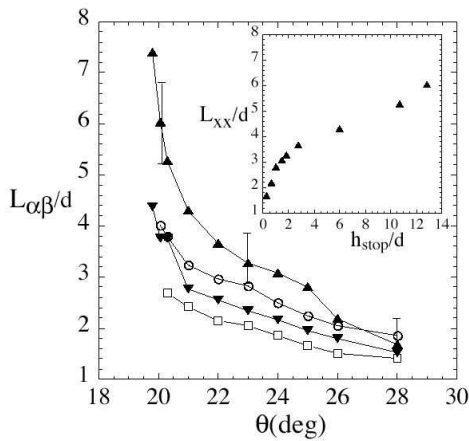}
  \caption{ The four correlation lengths $L_{xx}$ ($\blacktriangle$)  , $L_{yy}$ ({\Large $\circ$}) , $L_{yx}$ ($\blacktriangledown$), $L_{xy}$   ( $\square$)., as function of $\theta$; inset: $L_{xx}$ as a function of  $h_{stop}$.  The error bars give the measurement dispersion when the correlation lengths are computed from individual runs  and not from the mean correlation functions of Fig. \ref{fig6}.}
\label{fig7}
  \end{center}
 \end{figure}

The three plots in Fig.  \ref{fig5} present the normalized correlation function $C_{xx}/C_{xx}(0)$ at three different inclinations for different flow thicknesses $h$.  The striking result shown by these figures is the independence of the correlation function with the thickness $h$. At a given angle $\theta$, the normalized correlation function remains identical when one increases the flow thickness. This means that the spatial range of correlation is not controlled by the thickness but varies with inclination only. As a consequence, to each inclination corresponds a single correlation function, which can be computed by averaging the correlation functions obtained for different thicknesses.  The results obtained for all the inclinations are summarized in Fig. \ref{fig6}. The correlation function is steeper and steeper when increasing the plane inclination  $\theta$, showing  that the correlation length decreases. In order to quantify this effect, we define a correlation length $L_{xx}$ as the length where the correlation is equal to 0.07 times its value at the origin: $C_{xx}(L_{xx})=0.07 C_{xx}(0)$. Other definition gives similar qualitative results. Our measurements then show that $L_{xx}$ is a decreasing function of the inclination $\theta$ as shown in Fig. \ref{fig7}.  At high inclination, close to the maximum angle where steady uniform flows are observed, the correlation length is close to one particle diameter, which means that almost no correlation between neighboring particles exists. However, the correlation length dramatically increases for low inclinations when approaching the minimum angle at which a flow is possible. The same evolution with inclination, although less pronounced, is observed for the four correlation lengths $L_{\alpha \beta}$ (Fig. \ref{fig7}). 
This variation of the correlation length with inclination is reminiscent of the deposit function $h_{stop}(\theta)$ (Fig. \ref{fig1}). The deposit seems to diverge at the minimum angle where flow is possible and goes to zero at higher inclination.  Both quantities $L_{xx}$ and $h_{stop}$ thus varies in the same way with inclination as shown in  inset of Fig. \ref{fig7} where  $L_{xx}/d$ is plotted as a function $h_{stop}/d$ for the different inclinations. 


Given the error bars in the measurement of the correlation length, it is difficult to conclude about a precise relation between $L_{xx}$ and $h_{stop}$. However, despite the uncertainties, three major conclusions can be drawn:

i) Correlated motions exist at the free surface of a dense granular flow on a rough inclined plane.

ii) The spatial extend of the correlated motion is not controlled by the thickness of the flowing layer but is fixed by the inclination only. 

iii) The correlation length is maximum at low inclination and decreases at high inclination, in a similar way the minimum thickness for which flow is possible decreases with inclination. 

These conclusions only hold for the particles at the free surface. It would be interesting to know to which extend they still hold in the bulk of the flow, a study that could be possible in numerical simulations. The study also only concerns the generic case where the roughness of the plane is made of the same beads as in the bulk. However, a  recent study \cite{goujon03} shows that the flow properties varies when varying the roughness. An interesting question is how the correlation length would be also affected. Despite these open questions, 
this work gives a first experimental evidence to the proposition by Ertas and Halsey \cite{ertas02} that the flow threshold $h_{stop}(\theta)$ observed in the inclined plane configuration is the signature of  correlated motions.  More generally speaking, our results strongly support the idea that the rheology of granular material in the dense regime might be controlled by correlated motion of grains on a length scale larger than the particle diameter.  The microscopic origin of the spatial correlations and how to properly take them into account in writing constitutive laws, remain important challenges. 

 This work has benefited from fruitful discussions with B. Andreotti, O. Dauchot, Y. Forterre and P. Jop.

\end{document}